\documentclass[usenatbib]{mnras}

\usepackage{amssymb,amsmath}
\usepackage{graphicx}
\usepackage{xcolor}
\usepackage{multirow}
\usepackage[T1]{fontenc}
\usepackage{ae,aecompl}
\usepackage{newtxtext,newtxmath}

\def\oiii{[O~{\sc iii}]$\lambda4959,5007$\AA\ }
\def\o5007{[O~{\sc iii}]$\lambda5007$\AA\ }
\def\f3e{[O~{\sc iii}]$\lambda5007$\AA(E)}
\def\o3n{[O~{\sc iii}]$\lambda5007$\AA(N)}

\def\n6583{[N~{\sc ii}]$\lambda6583$\AA\ }
\def\c6717{[S~{\sc ii}]$\lambda6717$\AA\ }
\def\d6731{[S~{\sc ii}]$\lambda6731$\AA\ }

\def\a6300{[O~{\sc i}]$\lambda6300$\AA\ }
\def\b6363{[O~{\sc i}]$\lambda6363$\AA\ }

\def\obj{SDSS J1451+2709}

\title[\obj]{\obj~ a normal blue quasar but mis-classified as a HII galaxy in the BPT diagram by flux ratios 
of narrow emission lines}

\author[Zhang X. G.]{XueGuang Zhang$^{1}$
\thanks{Contact e-mail: \href{mailto:xgzhang@njnu.edu.cn}{xgzhang@njnu.edu.cn}}\\
$^{1}$School of Physics and Technology, Nanjing Normal University,
          No. 1, Wenyuan Road, Nanjing, 210023, P. R. China}

\pubyear{2021}

\begin{document}
\label{firstpage}
\pagerange{\pageref{firstpage}--\pageref{lastpage}}
\maketitle

\begin{abstract}   %%%241
	In the manuscript, we discuss properties of \obj, a normal blue quasar but mis-classified 
as a HII galaxy in the BPT diagram (called as a mis-classified quasar). The emission lines around 
H$\alpha$ and around H$\beta$ are well measured by different model functions with broad Balmer lines 
described by Gaussian or Lorentz functions, in the SDSS spectrum in 2007 and in the KPNO spectrum 
in 1990. After considering variations of broad emission lines, different model functions lead to 
different determined fluxes of narrow emission lines, but the different narrow emission line flux 
ratios lead the \obj~ as a HII galaxy in the BPT diagram. In order to explain the unique properties 
of the mis-classified quasar \obj~ in the BPT diagram, two methods are proposed, the starforming 
contributions and compressed NLRs with high electron densities near to critical densities of forbidden 
emission lines. Unfortunately, the two methods cannot be preferred in the \obj, further efforts are 
necessary to find the physical origin of the unique properties of the mis-classified quasar \obj~ 
in the BPT diagram. Meanwhile, there are not quite different long-term variabilities of \obj~ from 
the normal quasars. The mis-classified quasar \obj, an extremely unique case or a special case 
among the normal quasars, could provide further clues on the applications of BPT diagrams to the 
normal broad line AGN and to narrow emission line objects, indicating part of narrow emission line 
HII galaxies actually harbouring central AGN activities.
\end{abstract}


\section{Introduction}
                        
\begin{figure}
\centering\includegraphics[width = 8cm,height=5.5cm]{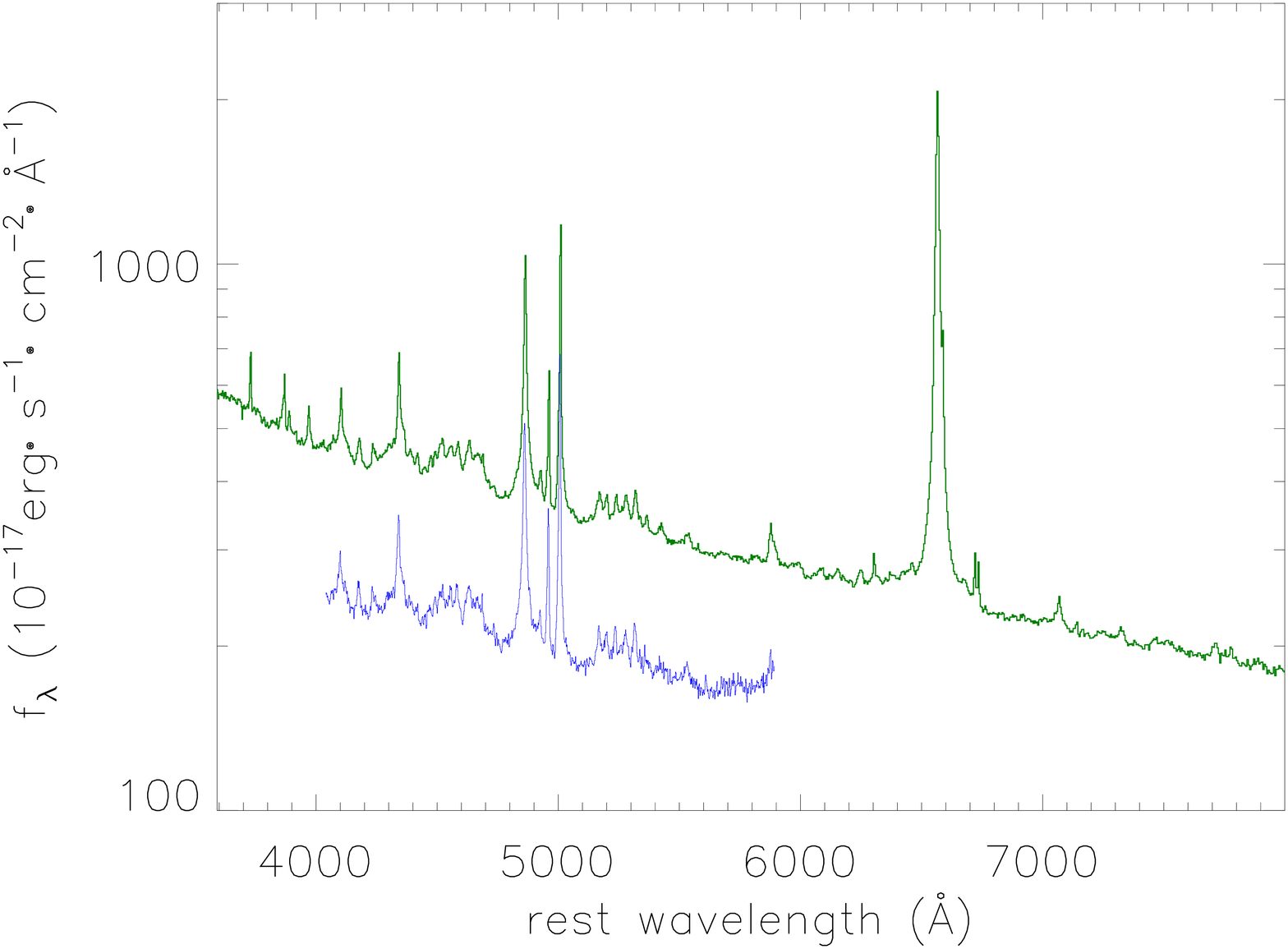}
\caption{The SDSS fiber spectrum (dark green) and the KNPO long-slit spectrum (blue) of \obj.
}
\label{spec}
\end{figure}

		%%%14:51:08.76, +27:09:26.93
	\obj~ (=~SDSS J145108+270926.92, also known as PG 1448+273) is always a common broad 
line blue quasar, since its first reported in the PG (Palomar-Green) quasar sample of \citet{bg92}. 
As the shown spectra in Fig.~\ref{spec} from SDSS (Sloan Digital Sky Survey) (the SDSS spectrum) and 
from the KPNO (Kitt Peak National Observatory) 2.1m telescope \citep{bg92} (the KPNO spectrum), 
\obj~ is an un-doubtfully normal broad line blue quasar (a type-1 AGN). However, based on flux 
ratios of narrow emission lines well discussed in Section 2 and in Section 3, \obj~ can be well 
classified as a HII galaxy instead of an AGN in the BPT diagram \citep{bpt81, kb01, ka03, kb06, 
ke13a, kb19, zh20}. Therefore, in the manuscript, some special properties of \obj~ is studied.

	Type-1 AGN (broad line Active Galactic Nuclei) and Type-2 AGN (narrow line AGN) having the 
similar properties of intrinsic broad and narrow emission lines, in the framework of the commonly 
accepted and well-known constantly being revised Unified Model \citep{an93, ra11, nh15, aa17, bb18, 
bn19, kw21} which has been successfully applied to explain different features between Type-1 and 
Type-2 AGN in many different ways. In other words, central broad line regions (BLRs) with distances 
of tens to hundreds of light-days \citep{kas00, kas05, bp06, bp09, dp10, ben13, fa17} to central 
black holes (BHs) are totally obscured by central dust torus (or other high density dust clouds), 
leading to no broad emission lines (especially in optical band) in observed spectra of Type-2 AGN. 
However, much extended narrow line regions (NLRs) with distances of hundreds to thousands of pcs 
(parsecs) to central BHs \citep{fi13, ha14, fi17, sg17} lead to the similar observed properties of 
narrow emission lines in Type-1 AGN and in Type-2 AGN, such as the strong linear correlation between 
the AGN power-law continuum luminosity and the \oiii line luminosity \citep{za03, sh11, hb14, zh17}. 
Type-2 AGN are intrinsically like Type-1 AGN expected by the Unified Model, strongly indicating that 
central AGN activities not only can be determined by narrow emission line properties but also can be 
determined by broad emission line properties.

	For an emission line object, there are two main methods applied to classify whether the 
object is an AGN or not? On the one hand, for a broad emission line object, through the clear 
spectroscopic features of broad emission lines and the strong blue power-low continuum emissions, 
a broad emission line object can be conveniently and directly classified as a type-1 AGN. On the 
other hand, for a narrow emission line object, the well-known BPT diagrams \citep{bpt81, ka03, 
ke06, ke13a,ke13, kb19, zh20} can be conveniently applied to determine whether the object is a 
Type-2 AGN (narrow line AGN) or a HII galaxy by properties of flux ratios of narrow emission lines. 
If there were similar intrinsic central activity properties, we could expect that the flux ratios 
of narrow emission lines for broad line objects (type-1 AGN) can also lead the broad line AGN to 
be well classified as AGN in the BPT diagrams. However, if there were some special bright type-1 
AGN (called as mis-classified AGN in the manuscript), of which flux ratios of narrow emission 
lines lead the broad line AGN classified as HII galaxies in the BPT diagrams, it would be very 
interesting to check properties of the special mis-classified AGN, which could probably provide 
further clues on unique properties of central emission structures.

	Not similar as standard narrow emission line AGN (type-2 AGN), there are more complicated 
line profiles of emission lines in broad line AGN, not only narrow emission components but also broad 
emission components in the spectrum. In some cases, the broad emission lines have line width (second 
moment) around several hundreds of kilometers per second, gently wider than extended components of 
forbidden/permitted emission lines, such as the commonly known extended components in 
[O~{\sc iii}]$\lambda4950,~5007$\AA~ doublet discussed in \citet{gh05, sh11, zh17, zh21a, zh21m}. It 
would be more difficult to determine narrow line flux ratios in some broad line AGN, until the narrow 
emission components can be well measured and confirmed. Therefore, variability properties of emission 
components in multi-epoch spectra of broad line AGN should be efficient to determine the emission regions 
in NLRs or in BLRs. 

	In the manuscript, among the SDSS pipeline classified low-redshift quasars ($z~<~0.3$) 
\citep{rg02, ro12, pr15, lh20}, \obj~ (PG 1448+273) is collected as the target, because of the following 
two main reasons. On the one hand, \obj~ has its type classified as a HII galaxy in the BPT diagram by 
flux ratios of narrow emission lines as well discussed in the following sections. On the other hand, \obj~ 
has two-epoch spectra observed in SDSS in 2007 and observed by the KPNO telescope in 1990, which will 
provide robust evidence to confirm the more reasonable flux ratios of narrow emission lines.

	The manuscript is organized as follows. Section 2 shows the properties of the spectroscopic 
emission lines by different model functions with different considerations. Section 3 shows the 
properties of \obj~ in the BPT diagram. Section 4 discusses the probable physical origin of the 
unique properties of the mis-classified quasar \obj~ in the BPT diagram. Section 5 shows the long-term 
variability properties of the mis-classified quasar \obj. Section 6 shows the further implications. 
Section 7 gives our final summaries and conclusions. And in the manuscript, the cosmological parameters 
of $H_{0}~=~70{\rm km\cdot s}^{-1}{\rm Mpc}^{-1}$, $\Omega_{\Lambda}~=~0.7$ and $\Omega_{m}~=~0.3$ 
have been adopted.

\begin{figure*}
\centering\includegraphics[width = 18cm,height=6cm]{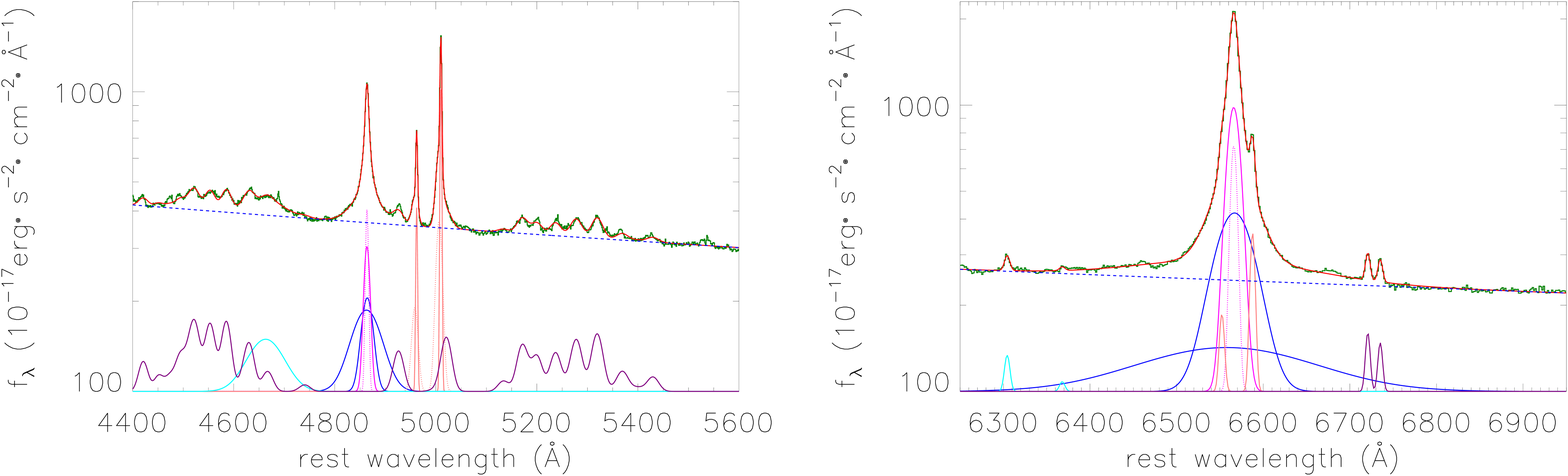}
\caption{The best fitting results to the emission lines around H$\beta$ (left panel) and H$\alpha$ 
(right panel) by multiple Gaussian functions plus power-law continuum emissions. In each panel, solid 
dark green line shows the SDSS spectrum, solid red line shows the determined best-fitting results, 
dashed blue line shows the determined power law component underneath the emission lines. In the left 
panel, solid blue lines show the determined broad H$\beta$ described by two broad Gaussian functions, 
solid purple line shows the determined optical Fe~{\sc ii} emission features, solid cyan line shows 
the determined broad He~{\sc ii} line, solid magenta line and dotted magenta line show the determined 
extended and core components in narrow H$\beta$, solid pink line and dotted pink line show the 
determined core and extended components in [O~{\sc iii}] doublet. In the right panel, solid blue 
lines show the determined two broad components in broad H$\alpha$ described by two broad Gaussian 
functions, solid magenta line and dotted magenta line show the determined extended and core components 
in narrow H$\alpha$, solid pink lines show the determined [N~{\sc ii}] doublet, solid cyan lines show 
the determined [O~{\sc i}] doublet, and solid purple lines show the determined [S~{\sc ii}] doublet. 
In order to show clear emission features, the Y-axis is in logarithmic coordinate in each panel.
}
\label{line}
\end{figure*}

\begin{figure*}
\centering\includegraphics[width = 18cm,height=6cm]{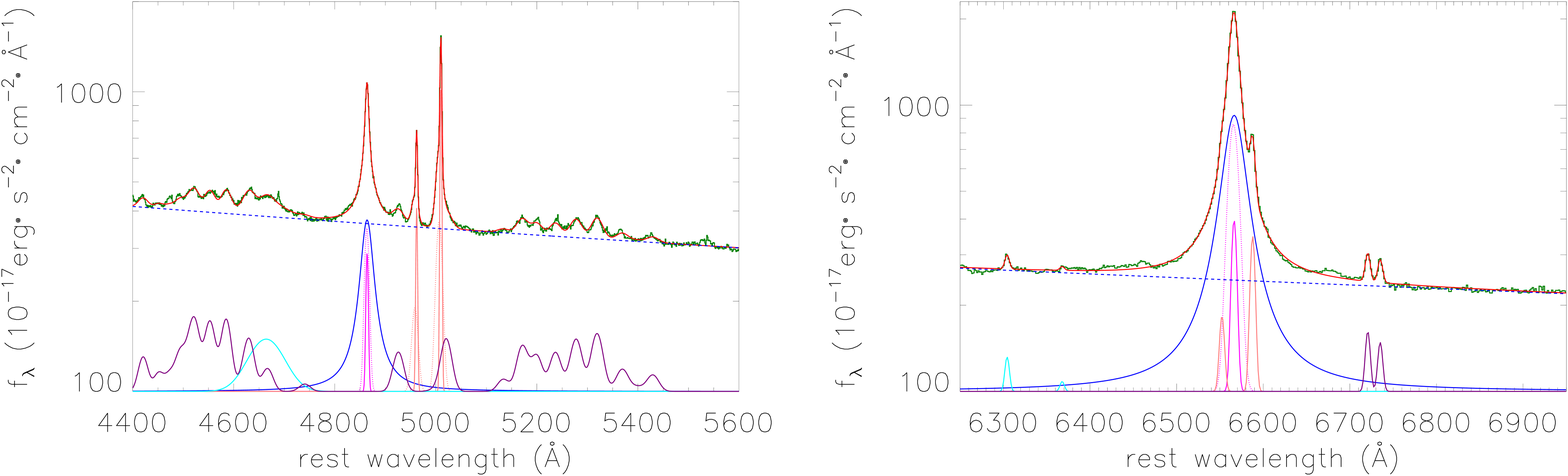}
\caption{Same as Fig.~\ref{line}, but the broad Balmer lines are described by Lorentz functions. 
In each panel, the solid blue line shows the Lorentz-function described broad Balmer line, the 
other line styles have the same meanings as those in Fig.~\ref{line}.
}
\label{llor}
\end{figure*}

\begin{figure}
\centering\includegraphics[width = 8cm,height=6cm]{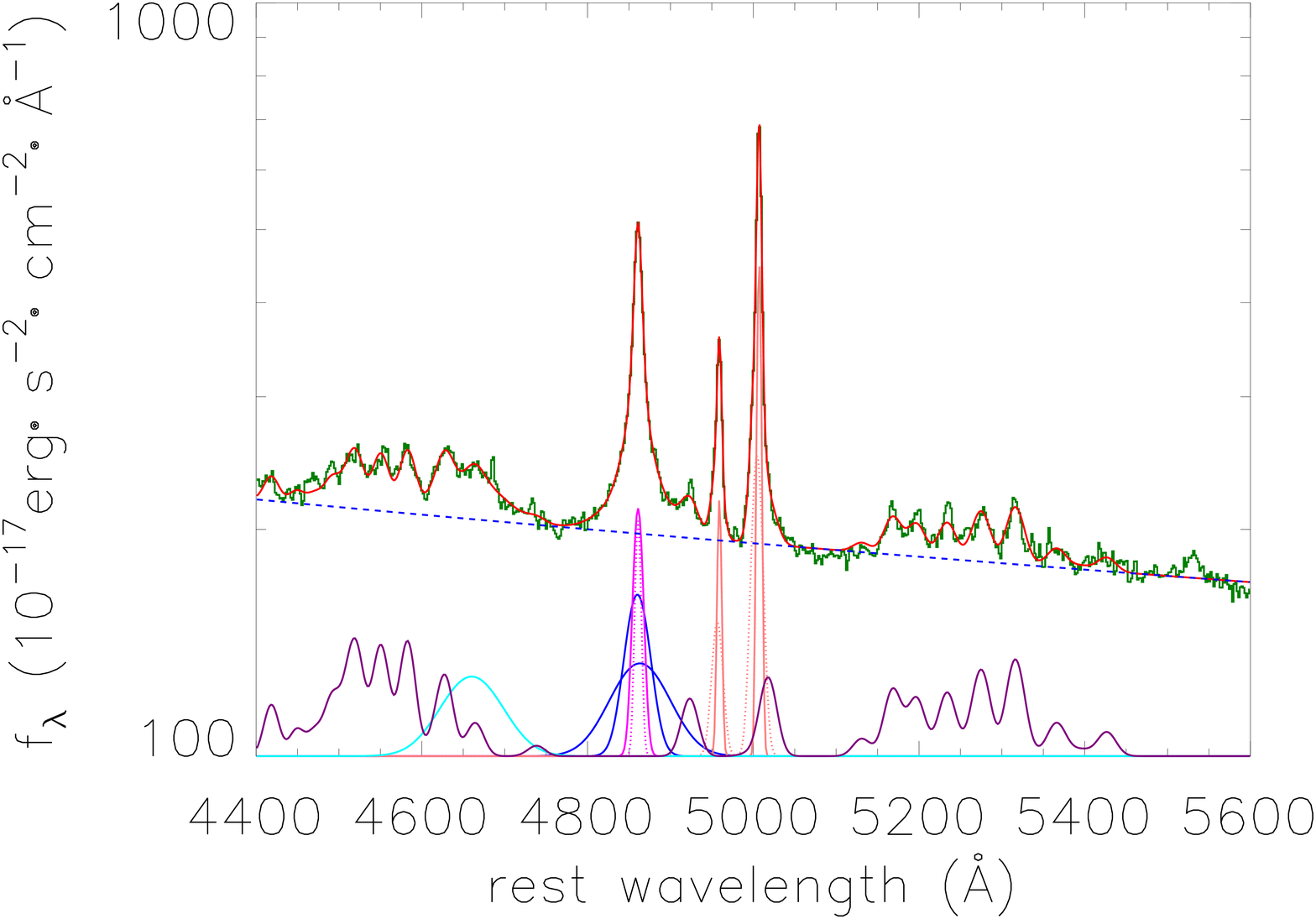}
\centering\includegraphics[width = 8cm,height=6cm]{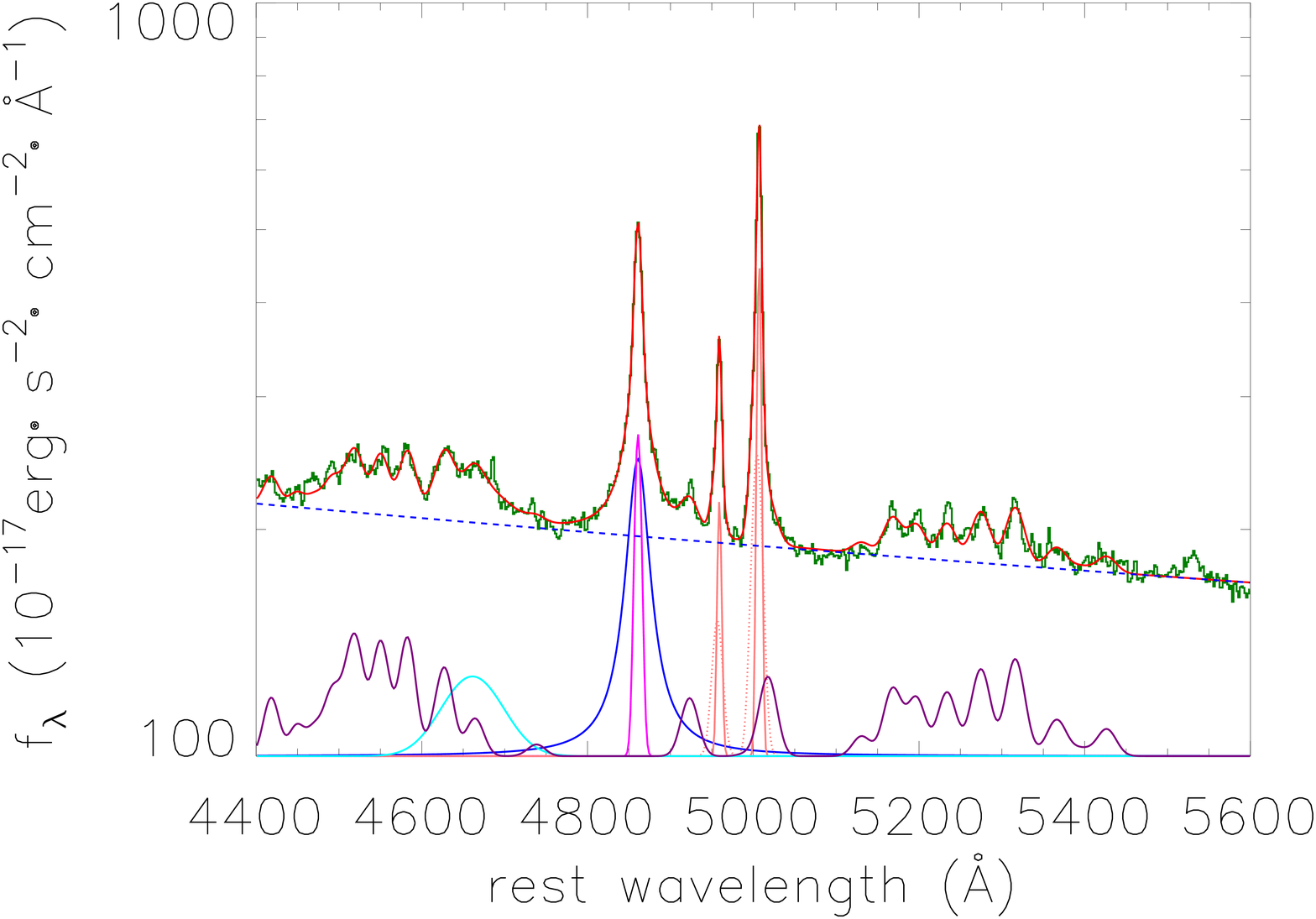}
\caption{Same as Fig.~\ref{line} and Fig~\ref{llor}, but for the emission lines around H$\beta$ in 
the KPNO long-slit spectrum in 1990. Top panel shows the results based on the broad H$\beta$ 
described by two broad Gaussian functions. Bottom panel shows the results based on the broad 
H$\beta$ described by a broad Lorentz function. The line styles have the same meanings as those 
in Fig.~\ref{line} and Fig~\ref{llor}
}
\label{bg92}
\end{figure}

%%%%%%%%plots are stopping here!

\section{Properties of spectroscopic emission lines of \obj}

	Fig.~\ref{spec} shows the SDSS spectrum of \obj~ observed in 2007 and the KPNO spectrum 
observed in 1990, with apparent broad emission lines and blue power-law continuum emissions. It 
is clear that \obj~ is an un-doubtfully broad line blue quasar. In order to show the classification 
of \obj~ by flux ratios of narrow emission lines in the BPT diagram, the following emission line 
fitting procedures with different model functions are applied to describe the emission lines of \obj, 
especially the emission lines of narrow H$\beta$, narrow H$\alpha$, [O~{\sc iii}]$\lambda4959,~5007$\AA~ 
doublet and [N~{\sc ii}]$\lambda6548,~6583$\AA~ doublet which will be applied in the following BPT 
diagram.

	Here, we mainly consider the emission lines around H$\beta$ and around H$\alpha$ (rest 
wavelength from 4400 to 5600\AA\ and from 6250 to 6950\AA), which are fitted simultaneously by the 
following two different kinds of model functions. The fitting procedure is very similar as what we 
have done in \citet{zh16, zh17, zh21a, zh21m}.

	For model A, the model functions are mainly based on Gaussian functions applied to describe 
the emission lines as follows. Two narrow plus two broad Gaussian functions applied to describe the 
core and extended components of [O~{\sc iii}]$\lambda4959,~5007$\AA~ doublet. Here, the broad Gaussian 
function means its second moment larger than the narrow Gaussian function, no restrictions on the 
lower limit of the second moment of the broad Gaussian function. One narrow and one broad Gaussian 
functions applied to describe the narrow H$\beta$. One narrow and one broad Gaussian functions applied 
to describe the narrow H$\alpha$. Two\footnote{More than two broad Gaussian functions have also been 
applied to describe the broad Balmer lines, however, the third or more broad Gaussian components are 
not necessary, because of the corresponding determined model parameters smaller than their uncertainties.} 
broad Gaussian functions are applied to describe the broad H$\beta$. Two broad Gaussian functions are 
applied to describe the broad H$\alpha$. Two\footnote{It is not necessary to consider extended 
components in [O~{\sc i}], [S~{\sc ii}] and [N~{\sc ii}] doublets. If additional Gaussian components 
are applied to describe probable extended components in the doublets, the determined line fluxes of 
the extended components are near to zero and smaller than determined uncertainties. Therefore, in the 
manuscript, besides the [O~{\sc iii}] doublet, there are no considerations on extended components 
of the forbidden emission lines.} narrow Gaussian functions are applied to describe the 
[O~{\sc i}]$\lambda6300,~6363$\AA~ doublet. Two narrow Gaussian functions are applied to describe the 
[S~{\sc ii}]$\lambda6716,~6731$\AA~ doublet. Two narrow Gaussian functions are applied to describe the 
[N~{\sc ii}]$\lambda6548,~6583$\AA~ doublet. One broad Gaussian function is applied to describe the 
broad He~{\sc ii} line. The broadened optical Fe~{\sc ii} template in \citet{kp10} is applied to 
describe the optical Fe~{\sc ii} emission features. One power law component is applied to describe 
the AGN continuum emissions underneath the emission lines around H$\beta$. One power law component is 
applied to describe the AGN continuum emissions underneath the emission lines around H$\alpha$. For 
the model parameters in model A, the following restrictions are accepted. First, the flux of each 
Gaussian component is not smaller than zero. Second, the flux ratio of the [O~{\sc iii}] doublet 
([N~{\sc ii}] doublet) is fixed to the theoretical value 3. Third, the Gaussian components of each 
forbidden line doublet have the same redshift and the same line width in velocity space. There are no 
further restrictions on the parameters of Balmer emission lines.

	For model B, the model functions are similar as the ones in model A, but the broad H$\beta$ 
(H$\alpha$) is described by only one broad Lorentz function. And the same restrictions are accepted 
to the model parameters in model B. The main objective to consider Lorentz function to describe 
the broad Balmer lines is as follows. Not similar as Gaussian function, Lorentz function always has 
sharp peak which can lead to more smaller measured fluxes of narrow Balmer lines, which will have 
positive effects on the classifications as HII galaxies by flux ratios of narrow emission lines in 
BPT diagram, which will be discussed in the following section. Here, we do not discuss the physical 
origin of Lorentz function described broad Balmer emission components, but the Lorentz function 
described broad Balmer emission components can lead to more reliable results on the \obj~ being 
classified as a HII galaxy in the BPT diagram.

	Through the Levenberg-Marquardt least-squares minimization technique, the emission lines 
around H$\beta$ and around H$\alpha$ can be well measured. The best fitting results are shown in 
Fig.~\ref{line} with $\chi^2/dof~=~0.79$ (summed squared residuals divided by degree of freedom) by 
model A, and in Fig~\ref{llor} with $\chi^2/dof~=~0.96$ by model B. The line parameters of each 
Gaussian component of narrow emission lines, Gaussian and Lorentz describe broad Balmer lines, 
and the power law continuum emissions are listed in Table~1.

	Based on the similar model functions in model A and in model B, but with no considerations 
on emission lines around H$\alpha$ (not covered in the KPNO spectrum), the emission lines around H$\beta$ 
in the KPNO long-slit spectrum in 1990 can be also well measured. The best fitting results are shown 
in the top panel of Fig.~\ref{bg92} with broad H$\beta$ described by two broad Gaussian functions, 
and in the bottom panel of Fig~\ref{bg92} with broad H$\beta$ described by one broad Lorentz function. 
The line parameters of each Gaussian component of narrow emission lines, Gaussian and Lorentz described 
broad H$\beta$ and the power law continuum emissions are also listed in Table~1. Here, due to loss of 
uncertainty information of the KPNO spectrum collected from NED (NASA$/$IPAC Extragalactic Database 
(\url{http://ned.ipac.caltech.edu/results/NEDspectra_output\_127\_page1_details.html\#PG\_1448+273\_2}), 
no uncertainties of determined model parameters are listed in Table~1 for the spectroscopic features 
in 1990.

\begin{table*}
\caption{Line parameters of the emission lines}
\begin{tabular}{lccl|ccl}
\hline\hline
name & $\lambda_0$ & $\sigma$ & flux &  $\lambda_0$ & $\sigma$ & flux  \\
\hline
	\multicolumn{7}{c}{Emission lines in the SDSS spectrum}\\
\hline
    &  \multicolumn{3}{c}{Gaussian Broad Balmer lines} & 
	\multicolumn{3}{c}{Lorentz Broad Balmer lines}  \\
    &  \multicolumn{3}{c}{$\chi^2/dof~=~0.79$} &  \multicolumn{3}{c}{$\chi^2/dof~=~0.96$} \\
\hline
H$\beta_{B1}$ & 4862.77$\pm$0.74 & 28.86$\pm$1.52 & 6.28$\pm$0.46 & 
	4863.89$\pm$0.19 & 29.15$\pm$1.36 & 12.52$\pm$0.22 \\ 
H$\beta_{B2}$ & 4864.27$\pm$0.48  & 11.21$\pm$1.65 & 2.96$\pm$0.45 & \dots & \dots & \dots \\
He~{\sc ii} & 4662.79$\pm$1.33 & 34.88$\pm$1.24 & 4.30$\pm$0.15 & 
	4664.24$\pm$1.33 & 34.51$\pm$1.21 & 4.28$\pm$0.15 \\
H$\beta_E$ & 4863.61$\pm$0.07 & 5.98$\pm$0.23 & 3.06$\pm$0.69 & 
	4863.39$\pm$0.13 & 4.38$\pm$0.25 & 2.75$\pm$0.43 \\
H$\beta_N$ & 4863.70$\pm$0.08 & 2.83$\pm$0.17 & 2.16$\pm$0.32 & 
	4863.96$\pm$0.17 & 2.21$\pm$0.29 & 1.06$\pm$0.35 \\
\o3n & 5010.04$\pm$0.03 & 1.64$\pm$0.03 & 3.81$\pm$0.08 &  
	5010.04$\pm$0.03 & 1.64$\pm$0.03 & 3.77$\pm$0.08 \\
\f3e & 5006.49$\pm$0.15 & 6.06$\pm$0.11 & 4.08$\pm$0.09 & 
	5006.53$\pm$0.14 & 6.02$\pm$0.11 & 4.05$\pm$0.09 \\
H$\alpha_{B2}$ & 6557.68$\pm$2.18 & 98.88$\pm$3.84 & 10.18$\pm$0.32 &
		6566.46$\pm$ 0.14 & 31.19$\pm$1.31 & 40.24$\pm$1.12 \\
H$\alpha_{B1}$ & 6566.82$\pm$0.25 & 23.18$\pm$0.45 & 18.61$\pm$0.52 & 
	\dots & \dots & \dots \\
H$\alpha_E$ & 6565.87$\pm$0.12 & 8.07$\pm$0.39 & 17.85$\pm$0.76 &  
	6565.57$\pm$0.18 & 5.91$\pm$0.34 & 11.22$\pm$1.13 \\
H$\alpha_N$ & 6565.99$\pm$0.09 & 3.83$\pm$0.19 & 6.01$\pm$1.02 & 
	6566.34$\pm$0.23 & 2.99$\pm$0.39 & 2.23$\pm$0.94 \\
\n6583 & 6587.92$\pm$0.12 & 2.83$\pm$0.12 & 1.83$\pm$0.09 &  
	6587.76$\pm$0.11 & 2.77$\pm$0.13 & 1.73$\pm$0.09 \\
\a6300 & 6304.37$\pm$0.33 & 3.04$\pm$0.34 & 0.26$\pm$0.03 & 
	6304.32$\pm$0.33 & 2.52$\pm$0.34 & 0.20$\pm$0.02 \\
\b6363 & 6367.88$\pm$0.35 & 3.07$\pm$0.36 & 0.06$\pm$0.02 & 
	6367.83$\pm$0.33 & 2.54$\pm$0.34 & 0.05$\pm$0.02 \\
\c6717 & 6720.71$\pm$0.13 & 2.61$\pm$0.13 & 0.39$\pm$0.02 & 
	6720.68$\pm$0.14 & 2.74$\pm$0.14 & 0.43$\pm$0.02 \\
\d6731 & 6735.09$\pm$0.14 & 2.61$\pm$0.14 & 0.31$\pm$0.02 & 
	6735.06$\pm$0.14 & 2.75$\pm$0.14 & 0.33$\pm$0.02 \\
pow H$\beta$ & \multicolumn{3}{c}{$f_\lambda~=~(343.27\pm0.44)(\frac{\lambda}{5100\textsc{\AA}})^{-1.37\pm0.01}$} & 
	\multicolumn{3}{c}{$f_\lambda~=~(341.11\pm0.47)(\frac{\lambda}{5100\textsc{\AA}})^{-1.32\pm0.01}$} \\
pow H$\alpha$ & \multicolumn{3}{c}{$f_\lambda~=~(244.41\pm0.55)(\frac{\lambda}{6563\textsc{\AA}})^{-1.79\pm0.04}$} & 
	\multicolumn{3}{c}{$f_\lambda~=~(245.08\pm0.32)(\frac{\lambda}{6563\textsc{\AA}})^{-1.93\pm0.03}$} \\
\hline
\multicolumn{7}{c}{Emission lines in the KPNO spectrum}\\
\hline
    &  \multicolumn{3}{c}{Gaussian Broad Balmer lines} &
	\multicolumn{3}{c}{Lorentz Broad Balmer lines}  \\
	\hline
H$\beta_{B1}$ & 4863.21 & 34.53 & 2.84 & 4860.81 & 30.69 & 7.09 \\
H$\beta_{B2}$ & 4860.12 & 15.03 & 2.36 & \dots & \dots & \dots \\
He~{\sc ii} & 4659.86 & 36.11 & 2.49 &
	4661.11 & 35.71 & 2.47 \\
H$\beta_E$ & 4860.97 & 6.24 & 1.56 & $\dots$ & $\dots$ & $\dots$ \\
H$\beta_N$ & 4860.88 & 3.73 & 1.13 & 4860.91 & 4.14 & 1.74 \\
\o3n & 5007.04 & 2.58 & 2.34 & 5007.05 & 2.56 & 2.32 \\
\f3e & 5004.46 & 6.46 & 2.39 & 5004.52 & 6.35 & 2.36 \\
pow H$\beta$ & \multicolumn{3}{c}{$f_\lambda~=~(187.76\pm0.52)(\frac{\lambda}{5100\textsc{\AA}})^{-1.05\pm0.03}$} &
        \multicolumn{3}{c}{$f_\lambda~=~(186.65\pm0.52)(\frac{\lambda}{5100\textsc{\AA}})^{-1.01\pm0.03}$} \\
\hline
\end{tabular}\\
{\bf Note:} The first column shows the information of emission component listed. 
[O~{\sc iii}]$\lambda5007$\AA(E) and [O~{\sc iii}]$\lambda5007$\AA(N) mean the extended and the 
core component of [O~{\sc iii}]$\lambda5007$\AA. H$\beta_E$ and H$\beta_N$ mean the extended and 
the core component of narrow H$\beta$, and H$\alpha_E$ and H$\alpha_N$ mean the extended and the 
core component of narrow H$\alpha$. The 'pow H$\beta$' and the 'pow H$\alpha$' mean the power law 
continuum emissions around H$\beta$ and around H$\alpha$, respectively. The second column to the 
fourth column show the line parameters of central wavelength in unit of \AA, second moment in 
unit of \AA~ and line flux in unit of $10^{-14}{\rm erg\cdot~s^{-1}\cdot~cm^{-2}}$ of the 
determine components by Model A with two broad Gaussian functions (H$\beta_{B1}$, H$\beta_{B2}$ 
and H$\alpha_{B1}$, H$\alpha_{B2}$) applied to describe the broad Balmer lines. The fifth column 
to the seventh column show the line parameters of the determine components by Model B with one 
broad Lorentz function (H$\beta_{B1}$, H$\alpha_{B1}$) applied to describe the broad Balmer lines.
\end{table*}

	It is clear that different model functions lead to quite different line parameters of 
narrow Balmer emission lines but similar line parameters of [O~{\sc iii}] and [N~{\sc ii}] 
doublets, which will lead to quite different flux ratios of [O~{\sc iii}] to narrow H$\beta$ 
and [N~{\sc ii}] to narrow H$\alpha$, Therefore, before to discuss properties of \obj~ in the 
BPT diagram, it is necessary to discuss which emission component of Balmer line emissions 
is actually from narrow emission regions.

\begin{figure*}
\centering\includegraphics[width = 15cm,height=12.5cm]{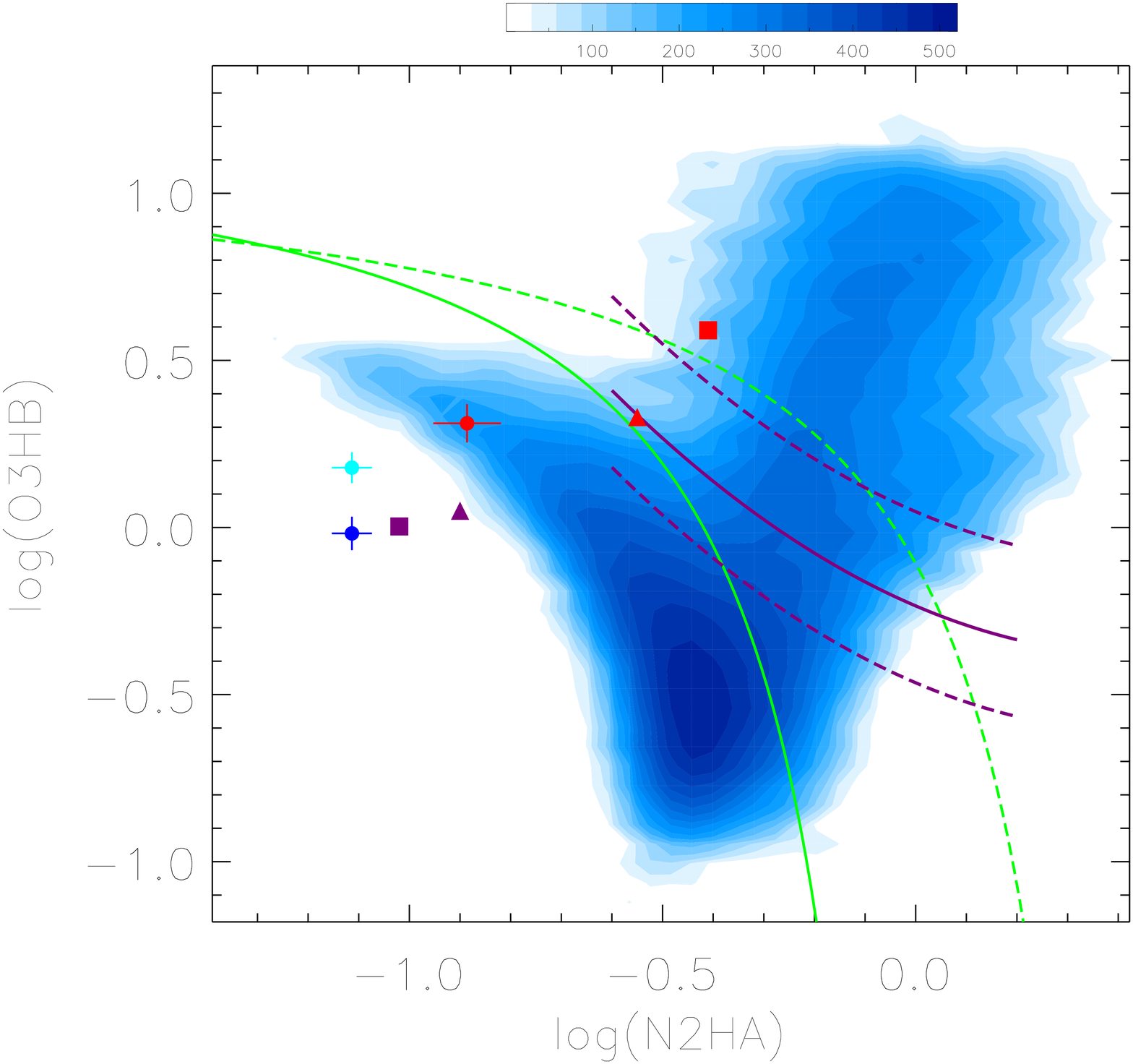}
\caption{The BPT diagram for more than 35000 narrow line objects (contour in dark green) and 
the mis-classified quasar \obj~ (solid circles in red, cyan, purple) by O3HB versus N2HA. Solid 
and dashed green lines show the dividing lines reported in \citet{ka03} and in \citet{kb01} 
between HII galaxies, composite galaxies and AGN. Solid purple line and dashed purple lines 
show the dividing line and area for composite galaxies determined in our recent work in 
\citet{zh20}. The contour is created by emission line properties of more than 35000 narrow 
emission-line galaxies discussed in \citet{zh20} collected from SDSS DR15. Corresponding 
number densities to different colors are shown in the color bar. Solid circles in blue, cyan 
and red represent the results of [$N2HA_A$, $O3HB_{LA}$], [$N2HA_A$, $O3HB_{UA}$] and 
[$N2HA_B$, $O3HB_{B}$], respectively.
}
\label{bpt}
\end{figure*}

\section{\obj~ in the BPT diagram}

	Before proceeding further, there are two points we should note. On the one hand, the 
SDSS spectrum is obtained by SDSS fibers with diameter of 3 arcseconds (about 4200pc), however 
the KPNO spectrum is obtained by a long-slit with aperture size of 1.5 arcseconds (about 2100pc). 
Therefore, different geometric emissions regions are covered in the SDSS spectrum and in the 
KPNO spectrum. However, two simple perspective can be reasonably accepted. First, the SDSS 
spectrum and the KPNO spectrum totally cover the AGN continuum emission regions. Second, the 
SDSS spectrum and the KPNO spectrum totally cover the central BLRs, due to the expected BLRs 
sizes about 16-23 light-days through the empirical relation between BLRs sizes and continuum 
luminosity well discussed in \citet{ben13}. Therefore, there should be apparent variabilities 
in the broad emission components from central BLRs from 1990 to 2007, because the continuum 
intensity at 5100\AA~ are brightened by 1.8 times. On the other hand, due to the loss of measured 
line parameters of the emission lines around H$\alpha$ in the KPNO spectrum, there are no 
further discussions of flux ratios of narrow emission lines in the KPNO spectrum.

\subsection{Flux ratios of narrow emission lines based on the model A}

	For the line parameters determined by model A with Gaussian functions applied to 
describe the broad Balmer lines, we can find that the broad Gaussian component H$\beta_{B1}$ 
in the broad H$\beta$ from 1990 to 2007 have line intensity brightened by 
$\frac{flux_{2007}}{flux_{1990}}~=~6.28/2.84$ times, meanwhile, have line width decreased 
by $\frac{\sigma_{2007}}{\sigma_{1990}}~=~34.53/28.86$. It is interesting to find that the 
variabilities of line width and line intensity of H$\beta_{B1}$ are well consistent with 
the Virialization assumption \citep{vm02, pe04, gh05b, vp06, rh11} expected result that 
\begin{equation}
(\frac{\sigma_{2007}}{\sigma_{1990}})^2~\sim~1.43~=~
	(\frac{flux_{2007}}{flux_{1990}})^{0.5}~\sim~1.48 
\end{equation}
Therefore, the broad emission component H$\beta_{B1}$ with line width larger than 
$1000{\rm km~s^{-1}}$ is from central BLRs.

	Unfortunately, there are no determined properties of emission components around 
H$\alpha$, because the KPNO spectrum only covers the wavelength range from 4030 to 5900\AA. 
However, considering the determined broad components H$\alpha_{B1}$ and H$\alpha_{B2}$ with 
line widths larger than 1000${\rm km\cdot~s^{-1}}$, the H$\alpha_{B1}$ and H$\alpha_{B2}$ 
can be safely accepted to be from central BLRs. Here, there is one point we should note. As 
discussed above, when the model A (model B) is applied to describe the emission lines, there 
are no restrictions on the line profiles of each broad component of broad Balmer lines, 
therefore, the determined H$\alpha_{B1}$ and H$\alpha_{B2}$ are quite different from the 
determined H$\beta_{B1}$ and H$\beta_{B2}$. 

%However, if to determine the central wavelength and second moment 
%of the broad Balmer lines including two broad Gaussian components, we can have
%\begin{equation}
%\lambda_0(H\alpha)~=~\frac{\int_{}^{}\lambda f_\lambda d\lambda}{\int\lambda d\lambda}~=~
%\end{equation}

	Besides the broad Gaussian component H$\beta_{B1}$ and H$\alpha_{B1}$ and H$\alpha_{B2}$ 
truly from central BLRs confirmed by the variability properties and by quite large line widths, 
the [O~{\sc iii}] doublet, both the core and the extended components, can be totally confirmed 
with emission regions in central NLRs. There is no doubt on the core components of the emission 
lines coming from central NLRs. Meanwhile, the extended component of [O~{\sc iii}] doublet with 
line width about $380{\rm km~\cdot~s^{-1}}$ can also be confirmed to be from central NLRs, based 
on the following main consideration. If the determined extended component of [O~{\sc iii}] doublet 
were not from [O~{\sc iii}] emission regions but from broad Balmer emission regions, there should 
be corresponding Gaussian components from Balmer emissions with rest wavelength about 6760\AA. 
However, the expected component can not be detected around H$\alpha$. Therefore, the determined 
extended components of [O~{\sc iii}] doublet are truly from central NLRs. Comparing with line 
widths of the core and the extended [O~{\sc iii}] components, the determined core and extended 
components of Balmer emissions, H$\beta_E$, H$\beta_N$, H$\alpha_E$, H$\alpha_N$, can be well 
accepted to come from central NLRs, because their line widths smaller than the line width of the 
extended component of [O~{\sc iii}] line. Certainly, the other forbidden line doublets are truly 
from central NLRs.

	Besides the emission components discussed above, the determined broad component H$\beta_{B2}$ 
has line width larger than the line width of the extended [O~{\sc iii}] component but smaller than 
$1000{\rm km\cdot~s^{-1}}$, and meanwhile, the broad component H$\beta_{B2}$ from 1990 to 2007 
have not apparent variabilities in line intensity. Although there are slightly increased line 
width in H$\beta_{B2}$ from 1990 to 2007, it is hard to confirm the broad component H$\beta_{B2}$ 
with line width about $700-900{\rm km\cdot~s^{-1}}$ is from central BLRs.

	Finally, for the determined components shown in Fig.~\ref{line} and with parameters listed 
in the second column to the fourth column in Table~1, the broad components H$\beta_{B1}$, H$\alpha_{B1}$, 
H$\alpha_{B2}$ are truly from central BLRs, the [N~{\sc ii}] doublet and the core and extended 
components of [O~{\sc iii}] doublet and narrow Balmer lines are from central NLRs. Therefore, 
considering the H$\beta_{B2}$ coming from central NLRs and the determined model parameters of the 
emission components in SDSS spectrum, the lower limit of flux ratio of [O~{\sc iii}]$\lambda5007$\AA~ 
(both the core and the extended component) to narrow H$\beta$ (including three components, H$\beta_{B2}$, 
H$\beta_E$ and H$\beta_N$) can be estimated as 
\begin{equation}
\begin{split}
O3HB_{LA}&~=~\frac{f_{[O~\textsc{iii}]\lambda5007\textsc{\AA}(N)}~+~
	f_{[O~\textsc{iii}]\lambda5007\textsc{\AA}(E)}}{f_{H\beta_N}
	~+~f_{H\beta_E}~+~f_{H\beta_{B2}}}\\
	&~=~ 0.96\pm0.11
\end{split}
\end{equation}
If considering the H$\beta_{B2}$ coming from central BLRs and the determined model parameters of 
the emission components in SDSS spectrum, the upper limit of flux ratio of [O~{\sc iii}]$\lambda5007$\AA~ 
(both the core and the extended component) to narrow H$\beta$ (including two components, H$\beta_E$ and 
H$\beta_N$) can be estimated as
\begin{equation}
\begin{split}
	O3HB_{UA}&~=~\frac{f_{[O~\textsc{iii}]\lambda5007\textsc{\AA}(N)}~+~
	f_{[O~\textsc{iii}]\lambda5007\textsc{\AA}(E)}}{f_{H\beta_N}~+~f_{H\beta_E}}\\
	&~=~ 1.51\pm0.16
\end{split}
\end{equation}
Considering the model parameters of [N~{\sc ii}]$\lambda6583$\AA~ and the core and extended 
components of narrow H$\alpha$ listed in Table~1, the flux ratio of [N~{\sc ii}]$\lambda6583$\AA~ 
to narrow H$\alpha$ (including two components, H$\alpha_E$ and H$\alpha_N$) can be estimated as
\begin{equation}
	N2HA_{A}~=~\frac{f_{[N~\textsc{ii}]\lambda6583\textsc{\AA})}}
	{f_{H\alpha_N}~+~f_{H\alpha_E}}
	~=~ 0.077\pm0.007
\end{equation}

	Based on the determined narrow emission line ratios by model A, the \obj~ is plotted in the 
BPT diagram of $O3HB$ (flux ratio of [O~{\sc iii}] to narrow H$\beta$) versus $N2HA$ (flux ratio of 
[N~{\sc ii}] to narrow H$\alpha$) in Fig.~\ref{bpt}. Considering the dividing lines in the BPT diagram 
as well discussed in \citet{ka03, kb06, kb19, zh20}, either [$N2HA_A$,~$O3HB_{UA}$] or [$N2HA_A$,
~$O3HB_{LA}$] applied in the BPT diagram through the properties of narrow emission lines, the \obj~ 
can be well classified as a HII galaxy with few contributions of central AGN activities, against the 
physical properties of a normal blue quasar.

\subsection{Flux ratios of narrow emission lines based on the model B}

	For the results by model B with broad Balmer lines described by Lorentz functions, 
the determined H$\beta_{B1}$ and H$\alpha_{B1}$ have line widths quite larger than 
$1000{\rm km\cdot s^{-1}}$, therefore, the determined H$\beta_{B1}$ and H$\alpha_{B1}$ 
can be safely accepted to be from central BLRs. However, the Lorentz described broad 
component H$\beta_{B1}$ have the same line widths about 30\AA~ ($1850{\rm km\cdot s^{-1}}$) 
in the KPNO spectrum in 1990 and in the SDSS spectrum in 2007. Meanwhile, as the results 
shown in Fig.~\ref{llor} and the parameters listed in Table~1, the continuum intensity 
at 5100\AA~ and the broad H$\beta$ flux are brightened by 1.8 times from 1990 to 2007, 
indicating line width of broad H$\beta$ in the SDSS spectrum in 2007 should be about 
1.16 times smaller than the one in the KPNO spectrum in 1990. However, the measured line 
width in 2007 is only 1.05times smaller than the value in 1990, indicating that the Lorentz 
described broad Balmer components include contributions of none-variability components 
(such as contributions from narrow Balmer emission lines from NLRs) or that the geometric 
structures of central BLRs are too extended.

	Certainly, the forbidden narrow lines are considered and well accepted from the 
central NLRs. Then, comparing with the line width $360{\rm km\cdot~s^{-1}}$ of extended 
component of [O~{\sc iii}]$\lambda$5007\AA, the determined components of narrow emission 
lines with line widths smaller than  $360{\rm km\cdot~s^{-1}}$ can be safely accepted 
to be from central NLRs.

	Finally, for the determined components shown in Fig.~\ref{llor} and with parameters 
listed in the fifth column to the seventh column in Table~1, the broad components H$\beta_{B1}$, 
H$\alpha_{B1}$ are truly from central BLRs, the [N~{\sc ii}] doublet and the core and 
extended components of [O~{\sc iii}] doublet and narrow Balmer lines are from central NLRs. 
Therefore, flux ratio of [O~{\sc iii}]$\lambda5007$\AA~ (both the core and the extended 
component) to narrow H$\beta$ (including two components, H$\beta_E$ and H$\beta_N$) can be 
estimated as
\begin{equation}
\begin{split}
O3HB_{B}&~=~\frac{f_{[O~\textsc{iii}]\lambda5007\textsc{\AA}(N)}~+~
	f_{[O~\textsc{iii}]\lambda5007\textsc{\AA}(E)}}{f_{H\beta_N}~+~f_{H\beta_E}} \\
	&~=~ 2.05\pm0.27
\end{split}
\end{equation}
Considering the model parameters of [N~{\sc ii}]$\lambda6583$\AA~ and the core and extended 
components of narrow H$\alpha$ listed in Table~1, the flux ratio of [N~{\sc ii}]$\lambda6583$\AA~ 
to narrow H$\alpha$ (including two components, H$\alpha_E$ and H$\alpha_N$) can be estimated as
\begin{equation}
N2HA_{B}~=~\frac{f_{[N~\textsc{ii}]\lambda6583\textsc{\AA})}}{f_{H\alpha_N}~+~f_{H\alpha_E}}
	~=~ 0.13\pm0.02
\end{equation}

	Based on the determined narrow emission line ratios by model B, the \obj~ can be 
re-plotted in the BPT diagram of $O3HB$ versus $N2HA$ in Fig.~\ref{bpt}. The \obj~ can be 
well classified as a HII galaxy with few contributions of central AGN activities, through 
the properties of narrow emission lines.

	Finally, based on different model functions to describe emission lines and based on 
different considerations of emission components from central NLRs, the \obj~ can be well 
classified as a HII galaxy with few contributions of central AGN activities. In the manuscript, 
the \obj~ can be called as a mis-classified quasar.

\section{Physical origin of the mis-classified quasar \obj?}

	In order to explain the mis-classified quasar \obj, two reasonable methods are mainly 
considered in the section. On the one hand, there is one mechanism leading to stronger narrow 
Balmer emissions, such as the strong starforming contributions. On the other hand, there is one 
mechanism leading to weaker forbidden emission lines, such as the expected high electron 
densities in central NLRs.

\subsection{Strong Starforming contributions?}

	In the subsection, we can check whether starforming contributions can be applied to 
explain the unique properties of \obj~ in the BPT diagram, because of stronger starforming 
contributions leading to stronger narrow Balmer emission lines. In other words, there are two 
kinds of flux components included in the narrow Balmer lines and [O~{\sc iii}]$\lambda5007$\AA~ 
and [N~{\sc ii}]$\lambda6583$\AA, one kind of flux depending on central AGN activities: 
$f_{[O~\textsc{iii}]}(AGN)$, $f_{[N~\textsc{ii}]}(AGN)$, $f_{H\alpha}(AGN)$ and 
$f_{H\beta}(AGN)$, the other kind of flux depending on starforming: $f_{[O~\textsc{iii}]}(SF)$, 
$f_{[N~\textsc{ii}]}(SF)$, $f_{H\alpha}(SF)$ and $f_{H\beta}(SF)$. Then, the measured flux 
ratio $O3HB$ and $N2HA$, and the flux rations $O3HB(AGN)$ and $N2HA(AGN)$ depending on 
central AGN activities, and the flux ratios $O3HB(SF)$ and $N2HA(SF)$ depending on starforming, 
can be described as  
\begin{equation}
\begin{split}
&O3HB~=~\frac{f_{[O~\textsc{iii}]}(AGN)~+~
	f_{[O~\textsc{iii}]}(SF)}{f_{H\beta}(AGN)~+~f_{H\beta}(SF)} \\
&O3HB(AGN)~=~\frac{f_{[O~\textsc{iii}]}(AGN)}{f_{H\beta}(AGN)} \\
&O3HB(SF)~=~\frac{f_{[O~\textsc{iii}]}(SF)}{f_{H\beta}(SF)} \\
&N2HA~=~\frac{f_{[N~\textsc{ii}]}(AGN)~+~
	f_{[N~\textsc{ii}]}(SF)}{f_{H\alpha}(AGN)~+~f_{H\alpha}(SF)} \\
&N2HA(AGN)~=~\frac{f_{[N~\textsc{ii}]}(AGN)}{f_{H\alpha}(AGN)} \\
&N2HA(SF)~=~\frac{f_{[N~\textsc{ii}]}(SF)}{f_{H\alpha}(SF)} \\
&f_{[O~\textsc{iii}]}~=~f_{[O~\textsc{iii}]}(AGN)~+~f_{[O~\textsc{iii}]}(SF) \\
&f_{H\beta}~=~f_{H\beta}(AGN)~+~f_{H\beta}(SF) \\
&f_{[N~\textsc{ii}]}~=~f_{[N~\textsc{ii}]}(AGN)~+~f_{[N~\textsc{ii}]}(SF) \\
&f_{H\alpha}~=~f_{H\alpha}(AGN)~+~f_{H\alpha}(SF)
\end{split}
\end{equation}
where $f_{[O~\textsc{iii}]}$, $f_{[N~\textsc{ii}]}$, $f_{H\alpha}$ and $f_{H\beta}$ mean 
the measured total line flux of [O~{\sc iii}]$\lambda5007$\AA, [N~{\sc ii}]$\lambda6583$\AA~ 
and narrow Balmer lines.

	Based on the three measured data points shown in Fig.~\ref{bpt} and the corresponding 
measured total line fluxes $f_{[O~\textsc{iii}]}$, $f_{[N~\textsc{ii}]}$, $f_{H\alpha}$ and 
$f_{H\beta}$, expected properties of starforming contributions $R_{SF}~=~f_{H\alpha}(SF)/f_{H\alpha}$ 
can be simply determined, through the following limitations. First, the determined flux ratios 
of $O3HB(AGN)$ and $N2HA(AGN)$ clearly lead the data points classified as AGN in the BPT diagram, 
the data points lying above the dividing line shown as solid green line in Fig.~\ref{bpt}. 
Second, the determined flux ratios of $O3HB(SF)$ and $N2HA(SF)$ clearly lead the data points 
classified as AGN in the BPT diagram, the data points lying below the dividing line shown as 
solid green line in Fig.~\ref{bpt}. Third, the ratios of $f_{H\alpha}(SF)$ to $f_{H\alpha}(AGN)$ 
are similar as the ratios of $f_{H\beta}(SF)$ to $f_{H\beta}(AGN)$.

\begin{figure*}
\centering\includegraphics[width = 18cm,height=10.5cm]{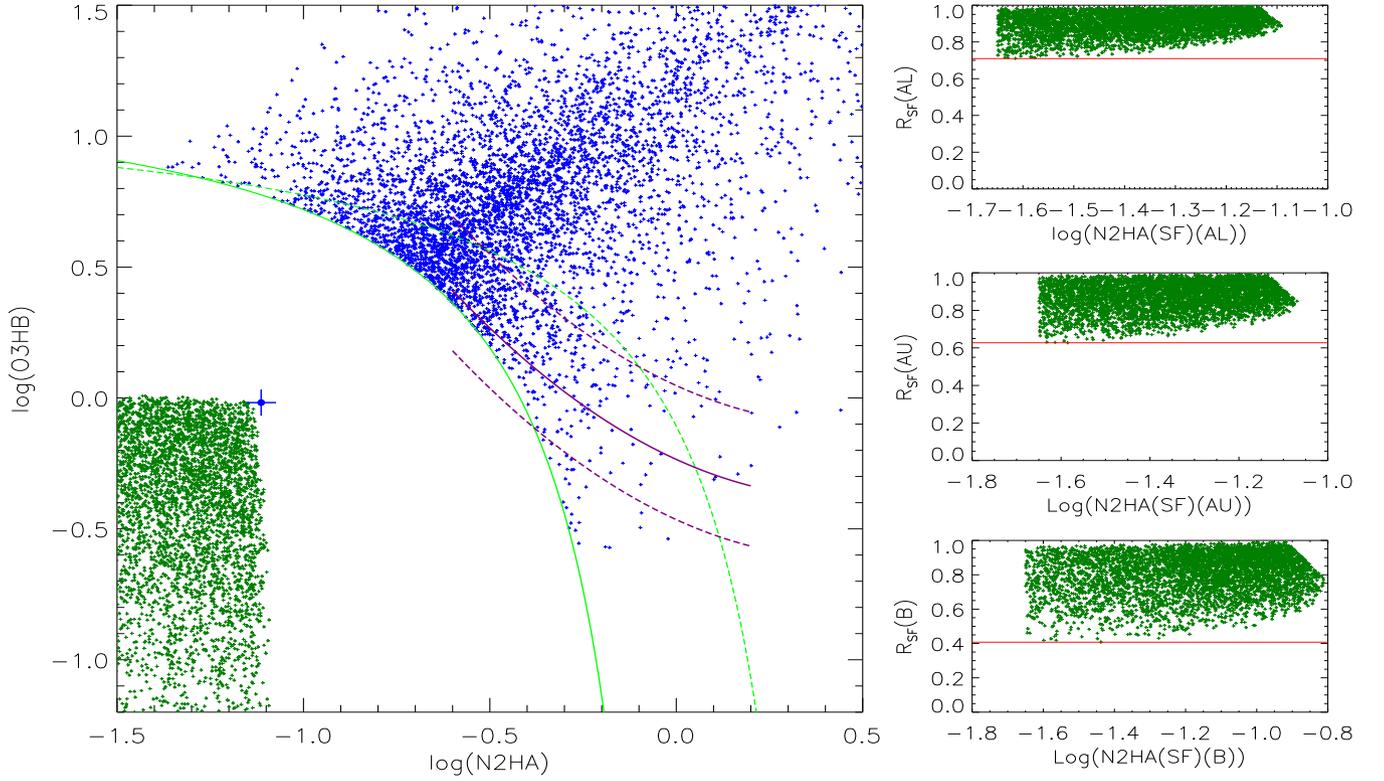}
\caption{Left panel shows the simulated 5000 couple data points of [$O3HB(AGN)$,~$N2HA(AGN)$] 
classified as AGN shown as blue pluses, and the simulated 5000 couple data points of 
[$O3HB(SF)$,~$N2HA(SF)$] classified as HII shown as dark green pluses, in the BPT diagram 
of $O3HB$ versus $N2HA$, based on the narrow line fluxes determined by model A with considering 
H$\beta_{B2}$ from central NLRs. The solid blue circle plus error bars are the same as those 
shown in Fig.~\ref{bpt}. Top right panel shows the dependence of $R_{SF}(AL)$ on $N2HA(SF)(AL)$, 
based on the narrow line fluxes determined by model A with considering H$\beta_{B2}$ from 
central NLRs. Middle right panel shows the dependence of $R_{SF}(AU)$ on $N2HA(SF)(AU)$, 
based on the narrow line fluxes determined by model A with considering H$\beta_{B2}$ from 
central BLRs. Bottom right panel shows the dependence of $R_{SF}(B)$ on $N2HA(SF)(B)$, based 
on the narrow line fluxes determined by model B. In each right panel, horizontal red line 
shows the position of the minimum value of $R_{SF}$. 
}
\label{cbpt}
\end{figure*}

	Based on the measured lime parameters by the model A with considering H$\beta_{B2}$ 
from NLRs, the narrow emission line fluxes are abut $f_{[O~\textsc{iii}]}~\sim~7.89$, 
$f_{[N~\textsc{ii}]}~\sim~1.83$, $f_{H\alpha}~\sim~23.86$ and $f_{H\beta}~\sim~8.18$ in the 
units of $10^{-14}{\rm erg\cdot~s^{-1}\cdot~cm^{-2}}$. Then, among 50000 randomly selected 
values of $f_{[O~\textsc{iii}]}(AGN)$, of $f_{[N~\textsc{ii}]}(AGN)$ and of $f_{H\alpha}(AGN)$, 
there are 5000 couple data points of [$O3HB(AGN)$,~$N2HA(AGN)$] classified as AGN, and 
corresponding 5000 couple data points of [$O3HB(SF)$,~$N2HA(SF)$] classified as HII, in the 
BPT diagram of $O3HB$ versus $N2HA$, shown in the left panel of Fig.~\ref{cbpt}. And top right 
panel of Fig.~\ref{cbpt} shows the dependence of $R_{SF}(AL)$ on $N2HA(SF)(AL)$, with the 
determined minimum value 71\% of the $R_{SF}(AL)$.

	Similarly, based on the measured lime parameters by the model A with considering 
H$\beta_{B2}$ from BLRs, the narrow emission line fluxes are abut $f_{[O~\textsc{iii}]}~\sim~7.89$, 
$f_{[N~\textsc{ii}]}~\sim~1.83$, $f_{H\alpha}~\sim~23.86$ and $f_{H\beta}~\sim~5.22$ in the 
units of $10^{-14}{\rm erg\cdot~s^{-1}\cdot~cm^{-2}}$. Then, among 40000 randomly selected 
values of $f_{[O~\textsc{iii}]}(AGN)$, of $f_{[N~\textsc{ii}]}(AGN)$ and of $f_{H\alpha}(AGN)$, 
there are 5000 couple data points of [$O3HB(AGN)$,~$N2HA(AGN)$] classified as AGN, and 
corresponding 5000 couple data points of [$O3HB(SF)$,~$N2HA(SF)$] classified as HII, in the 
BPT diagram of $O3HB$ versus $N2HA$. Here, we do not show the results in the BPT diagram, which 
are similar as the those shown in the left panel of Fig.~\ref{cbpt} but with different positions 
of the data points. And the middle right panel of Fig.~\ref{cbpt} shows the dependence of 
$R_{SF}(AU)$ on $N2HA(SF)(AU)$, with the determined minimum value 63\% of the $R_{SF}(AU)$.

	Based on the measured lime parameters by the model B, the narrow emission line fluxes 
are abut $f_{[O~\textsc{iii}]}~\sim~7.66$, $f_{[N~\textsc{ii}]}~\sim~1.73$, $f_{H\alpha}~\sim~13.45$ 
and $f_{H\beta}~\sim~3.81$ in the units of $10^{-14}{\rm erg\cdot~s^{-1}\cdot~cm^{-2}}$. Then, 
among 20000 randomly selected values of $f_{[O~\textsc{iii}]}(AGN)$, of $f_{[N~\textsc{ii}]}(AGN)$ 
and of $f_{H\alpha}(AGN)$, there are 5000 couple data points of [$O3HB(AGN)$,~$N2HA(AGN)$] 
classified as AGN, and corresponding 5000 couple data points of [$O3HB(SF)$,~$N2HA(SF)$] 
classified as HII, in the BPT diagram of $O3HB$ versus $N2HA$. Here, we do not show the results 
in the BPT diagram, which are similar as the those shown in the left panel of Fig.~\ref{cbpt} 
but with different positions of the data points. And the bottom right panel of Fig.~\ref{cbpt} 
shows the dependence of $R_{SF}(B)$ on $N2HA(SF)(B)$, with the determined minimum value 42\% 
of the $R_{SF}(B)$.

\begin{figure*}
\centering\includegraphics[width = 18cm,height=6cm]{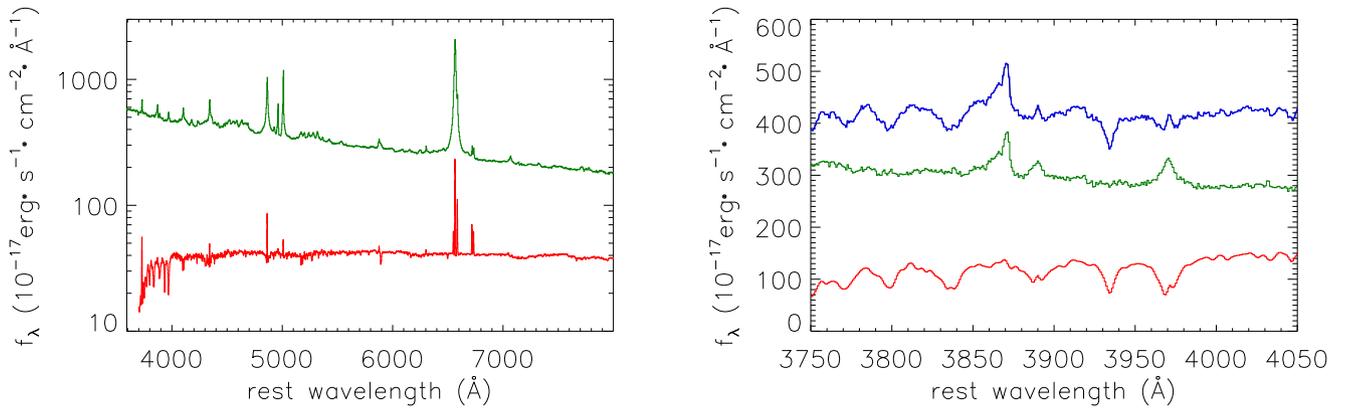}
\caption{Left panel shows the SDSS spectrum of \obj~ (in dark green) and the mean spectrum 
of HII galaxies (in red). Right panel shows the composite spectrum around 4000\AA, including 
40\% starforming contributions. In right panel, solid blue line shows the composite spectrum, 
solid dark green line shows the SDSS spectrum, and solid red line shows the starforming 
contributions.
}
\label{sp2}
\end{figure*}

	Therefore, by different model functions determined narrow line fluxes with different 
considerations, the determined starforming contributions $R_{SF}$ should be larger than 40\%, 
strongly indicating apparent absorption features from host galaxy in the spectrum of \obj. 
Fig.~\ref{sp2} shows one composite spectrum created by 0.6 times of the SDSS spectrum of \obj~ 
plus a mean HII galaxy with continuum intensity at 5100\AA~ about 0.4 times of the continuum 
intensity at 5100\AA~ of the SDSS spectrum of \obj. Here, the mean spectrum of HII galaxies 
are created by the large sample of 1298 HII galaxies with signal-to-noise larger than 30\ 
in SDSS DR12. The absorption features are apparent enough in the composite spectrum with 40\% 
starforming contributions. However, the clean quasar-like spectrum without clear absorption 
features, especially around 4000\AA, clearly indicate that the starforming contributions cannot 
be applied to explain the unique properties of the mis-classified quasar \obj~ in the BPT diagram.

\subsection{Compressed central NLRs?}

	Besides the starforming contributions, high electron density in NLRs can be also 
applied to explain the unique properties of the mis-classified quasar \obj~ in the BPT diagram, 
because the high electron density near to the critical electron densities of the forbidden 
emission lines can lead to suppressed line intensities of forbidden emission lines but 
positive effects on strengthened Balmer emission lines.

	It is not hard to determine electron density in NLRs, such as through the flux 
ratios of [S~{\sc ii}]$\lambda6717,6731$\AA~ doublet as well discussed in \citet{po14, 
sa16, kn19}. Based on the measured line fluxes of [S~{\sc ii}] doublet listed in Table~1, 
the flux ratio of [S~{\sc ii}]$\lambda6716$\AA~ to [S~{\sc ii}]$\lambda6731$\AA~ lead the 
electron density to be estimated around ${\rm 200cm^{-3}}$, a quite normal value, quite 
smaller than the critical densities around ${\rm 10^5 cm^{-3}}$ to [O~{\sc iii}] and 
[N~{\sc ii}] doublet.

	Based on the estimated electron density in NLRs, the compressed central NLRs with 
higher electron densities could not be preferred to explain the unique properties of the 
mis-classified quasar \obj~ in the BPT diagram. Certainly, the probability of the compressed 
central NLRs cannot be totally ruled out, considering different emission regions of [S~{\sc ii}] 
doublet from the emission regions of [O~{\sc iii}] and [N~{\sc ii}] doublet.

	Unfortunately, neither the starforming contributions nor the compressed NLRs can 
be preferred in the mis-classified quasar \obj. Further efforts are necessary to determine 
the physical origin of the unique properties of the mis-classified quasar \obj~ in the BPT 
diagram.

\section{Long-term photometric variability properties}

	In the section, long-term variabilities of \obj~ are well checked, in order to 
check whether are there further clues on unique variability properties of the mis-classified 
quasar \obj.

\begin{figure*}
\centering\includegraphics[width = 18cm,height=12cm]{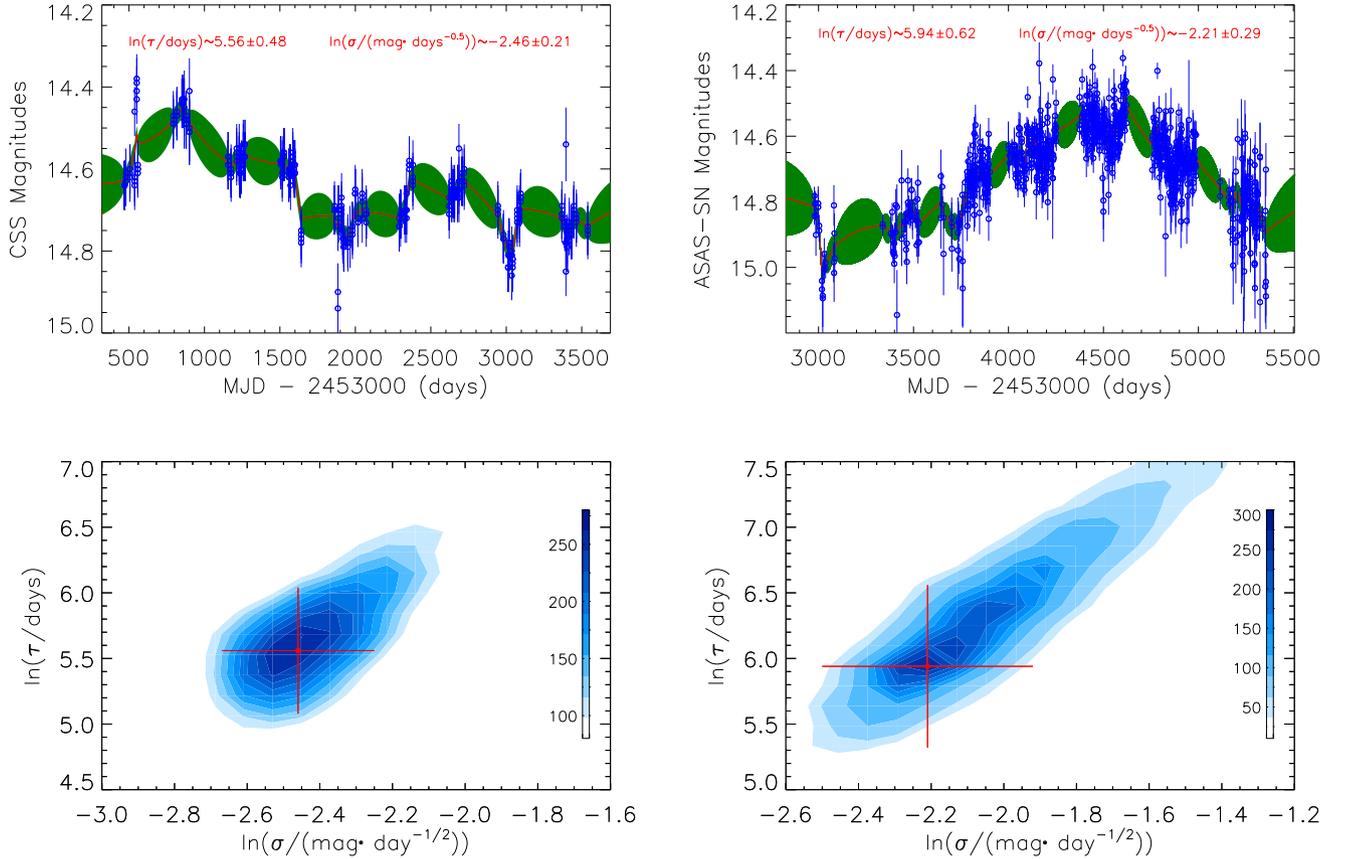}
\caption{Left panels show the photometric variability properties from the CSS. In top left 
panel, open circles plus error bars in blue show the photometric data points, solid red line 
and the regions filled by dark green show the DRW method determined best descriptions and 
the corresponding 1sigma confidence bands. Bottom left panel shows the posterior distributions 
in contour of the DRW process parameters of $\sigma$ and $\tau$ through the CSS light curve. 
The sold circle plus error bars in red in bottom left panel show the determined values and 
uncertainties of $\sigma$ and $\tau$ which are marked in red characters in top corner in the 
top left panel. Right panels shows the similar results, but for the photometric light curve 
from ASAS-SN.
}
\label{lmc}
\end{figure*}

	Besides the spectroscopic results for \obj, long-term photometric variability 
can be collected from the Catalina Sky Survey (CSS, \url{http://nesssi.cacr.caltech.edu/}) 
\citep{dra09, lar03} with with MJD from 2453470 to 2456567 shown in the top left panel 
in Fig.~\ref{lmc}, and from the well-known All-Sky Automated Survey for Supernovae 
(ASAS-SN, \url{https://asas-sn.osu.edu/}) \citep{sp14, ks17} with MJD from 2455978 to 2458737 
shown in the top right panel in Fig.~\ref{lmc}.

	Now it is interesting to check the long-term variability of the mis-classified quasar 
\obj~ by the well-known Damped Random Walk process (DRW process or Continuous AutoRegressive 
process, CAR process) \citep{bd02}. The DRW process, with two basic parameters of the intrinsic 
variability timescale $\tau$ and the intrinsic variability amplitude $\sigma$, has been proved 
to be a preferred modeling process to describe AGN intrinsic variability. \citet{kbs09} firstly 
proposed the CAR process to describe the AGN intrinsic variability, and found that the AGN 
intrinsic variability timescales are consistent with disk orbital or thermal timescales. 
\citet{koz10} provided an improved robust mathematic method to estimate the DRW process 
parameters, and found that AGN variability could be well modeled by the DRW process. Then, 
\citet{zu11} provided a public code of JAVELIN 
(\url{http://www.astronomy.ohio-state.edu/~yingzu/codes.html\#javelin}) (Just Another Vehicle for 
Estimating Lags In Nuclei) based on the method in \citet{koz10} to describe the AGN variability 
by the DRW process.

	Meanwhile, there are many other reported studies on the AGN variability through the 
DRW process. \citet{mi10} modeled the variability of about 9000 spectroscopically confirmed 
quasars covered in the SDSS Stripe82 region, and found correlations between the AGN parameters 
and the DRW process determined parameters. \citet{bj12} proposed an another fully probabilistic 
method for modeling AGN variability by the DRW process. \citet{ak13} have shown that the DRW 
process is preferred to model AGN variability, rather than several other stochastic and 
deterministic models, by fitted results of long-term variability of 6304 quasars. \citet{zu13} 
have checked that the DRW process provided an adequate description of AGN optical variability 
across all timescales. More recently, in our previous paper, \citet{zh17a} have checked long-term 
variability  properties of AGN with double-peaked broad emission lines, and found the difference 
in intrinsic variability timescales between normal broad line AGN and the AGN with double-peaked 
broad emission lines. Therefore, the DRW process determined parameters from the long-term AGN 
variability can be well used to check or predict further different properties of different kinds 
of AGN with probable different intrinsic properties of emission regions.

	The public code of JAVELIN provided by \citet{zu11, zu13} has been applied here to 
describe the long-term variability shown in top panels of Fig.~\ref{lmc}. When the JAVELIN 
code is applied, through the MCMC (Markov Chain Monte Carlo) analysis with the uniform logarithmic 
priors of the DRW process parameters of $\tau$ and $\sigma$ covering every possible corner of 
the parameter space ($0~{\rm days}<~\tau~<~1e+5~{\rm days}$ and 
$0~{\rm mag/day^{0.5}}<~\sigma~<~1e+2~{\rm mag\cdot~day^{-0.5}}$), the posterior distributions of 
the DRW process parameters can be well determined and shown in bottom panels of Fig.~\ref{lmc} 
and provide the final accepted parameters and the corresponding uncertainties. For the 
photometric CSS light curve, the determined DRW process parameters are 
$\ln(\tau/{\rm days})~\sim~5.56\pm0.48$ and $\ln(\sigma/(\rm mag\cdot~days^{-0.5}))~\sim~-2.46\pm0.21$. 
For the photometric ASAS-SN light curve, the determined DRW process parameters are
$\ln(\tau/{\rm days})~\sim~5.94\pm0.62$ and $\ln(\sigma/(\rm mag\cdot~days^{-0.5}))~\sim~-2.21\pm0.29$.
Although there are different time-steps and time durations of the CSS light curve and the 
ASAS-SN light curve, there are similar determined DRW process parameters of $\sigma$ and $\tau$, 
indicating the determined DRW process parameters are stable enough.

	Comparing with variabilities of quasars discussed in \citet{kbs09, mi10, koz10}, the 
mis-classified quasar \obj~ has the normal intrinsic variability timescale $\tau$ around 
300days, however, have a large but also acceptable intrinsic variability amplitude $\sigma$ 
around 0.1$\rm mag\cdot~days^{-0.5}$ (especially comparing with the results shown in 
Fig.~6\ in \citet{koz10}). It is clear that \obj~ has common long-term photometric variability 
properties.

\section{Further implications of the mis-classified quasar \obj}

	If the discussed results above are true for the mis-classified quasar \obj, there are 
at least three further implications on the study of classifications by the BPT diagrams.

	First and foremost, if the mis-classified quasar \obj~ was an extremely unique case 
among the quasars in the BPT diagram, there must be extremely special emission structures in 
\obj, leading to quite lower flux ratios of narrow emission lines than normal quasars. The 
extremely unique case of the mis-classified quasar J1451+2709 deserves further study in the 
near future.

	Besides, if the mis-classified quasar \obj~ was not an extremely unique case but a 
some special case among quasars in the BPT diagrams, there would be more mis-classified 
quasars which can be detected among normal quasars. To detect a sample of mis-classified 
quasars should provide clues that the mis-classified quasars should be a special subclass 
of type-1 AGN with some unknown special emission properties. Meanwhile, the results should 
indicate that there could be some narrow emission lines HII galaxies with probable central 
AGN activities.

	Last but not the least, if the mis-classified quasar \obj~ was due to mis-applications 
of BPT diagrams to broad line AGN, it would be a challenge to the applications of BPT diagrams 
from narrow emission line objects to broad line AGN. Or, emission line fitting procedure 
determined emission line parameters are not so efficient for narrow emission lines?

\section{Summaries and Conclusions}

	Finally, we give our main conclusions as follows. 
\begin{itemize}
\item Emission line parameters of the blue quasar \obj~ can be well measured by two different 
	models, mode A with broad Gaussian functions applied to describe broad Balmer emission 
	lines, and model B with broad Lorentz functions applied to described the broad Balmer 
	emission lines.
\item Based on two-epoch spectra observed in SDSS in 2007 and observed by KPNO in 1990, broad 
	Balmer emission components from central BLRs can be well determined. Besides the emission 
	components from central BLRs, emission components from central NLRs can be estimated, 
	leading to different flux ratios of narrow emission lines.
\item Different flux ratios of narrow emission lines determined by different model functions 
	and with different considerations, the \obj~ can be well classified as a HII galaxy in 
	the BPT diagram, although the \obj~ actually is a normal blue quasar.
\item Two reasonable methods are proposed to explain the unique properties of the mis-classified 
	quasar \obj~ in the BPT diagram, strong starforming contributions leading to stronger 
	narrow Balmer emissions, and compressed NLRs with high electron densities leading to 
	suppressed forbidden emissions. 
\item Once considering the starforming contributions, at least 40\% starforming contributions 
	should be preferred in the mis-classified quasar \obj, which will lead to apparent 
	absorption features around 4000\AA. However, none apparent absorption features in the 
	SDSS spectrum of the mis-classified quasar \obj~ indicate strong starforming contributions 
	are not preferred in the mis-classified quasar \obj.
\item Once considering the compressed NLRs with high electron densities, the expected electron 
	densities should be around ${\rm 10^5cm^{-3}}$. However, the estimated electron density 
	is only around ${\rm 200cm^{-3}}$ based on the flux ratio of [S~{\sc ii}]$\lambda6716$\AA~ 
	to [S~{\sc ii}]$\lambda6731$\AA. Therefore, the compressed NLRs with high electron densities 
	are not preferred in the mis-classified quasar \obj.
\item Further efforts are necessary to find reasonable physical origin of the unique properties 
	of the mis-classified quasar \obj~ in the BPT diagram.
\item The long-term variability properties of the mis-classified quasar \obj~ are not quite different 
	from the normal quasars.
\item The reported mis-classified quasar \obj~ strongly indicate that there should be extremely 
	unique properties of \obj~ which are currently not detected, or indicate that there should 
	be a small sample of mis-classified quasars similar as the \obj, or indicate that there 
	should be some narrow emission line HII galaxies intrinsically harbouring central AGN 
	activities.
\end{itemize}

\section*{Acknowledgements}
%{\color{red}Zhang gratefully acknowledge the anonymous referee for 
%giving us constructive comments and suggestions to greatly improve our paper.} 
Zhang gratefully acknowledges the anonymous referee reading the manuscript with patience. 
Zhang gratefully acknowledges the kind support of Starting Research Fund of Nanjing Normal 
University, and gratefully acknowledges the kind support of NSFC-11973029. This manuscript 
has made use of the data from the SDSS projects. The SDSS-III web site is http://www.sdss3.org/. 
SDSS-III is managed by the Astrophysical Research Consortium for the Participating Institutions 
of the SDSS-III Collaborations. The manuscript has made use of the data from from the Catalina 
Sky Survey \url{http://nesssi.cacr.caltech.edu/DataRelease/}, and from the All-Sky Automated 
Survey for Supernovae \url{https://asas-sn.osu.edu/}. The manuscript has made use of the 
NASA/IPAC Extragalactic Database (NED) \url{https://ned.ipac.caltech.edu/}, which is funded 
by the National Aeronautics and Space Administration and operated by the California Institute 
of Technology.

\section*{Data Availability}
The data underlying this article will be shared on reasonable request to the corresponding author
(\href{mailto:xgzhang@njnu.edu.cn}{xgzhang@njnu.edu.cn}).

\label{lastpage}
\end{document}